\documentclass[preprint,aps,showkeys]{revtex4-1}
\usepackage{amsmath}%
\usepackage{graphicx}%
\usepackage{color}%
\usepackage{dcolumn}%
\usepackage{bm}%
\usepackage{ulem}
\usepackage[uline]{hhtensor}
\usepackage{setspace}
\usepackage{siunitx}
\makeatletter
\renewcommand{\fnum@figure}{Figure \thefigure}
\makeatother

\begin{document}

\title{Effect of strain on interactions of $\Sigma$3\{111\} silicon grain boundary with Oxygen impurities from first-principles}

\author{Rita Maji}
\address{Dipartimento di Scienze e Metodi dell'Ingegneria, Universit{\`a} di Modena e Reggio Emilia,Via Amendola 2 Padiglione Tamburini , I-42122 Reggio Emilia, Italy}

\author{Julia Contreras-Garc\'{\i}a} 
\address{Laboratoire de Chimie Th\'eorique, Sorbonne Universit\'e and CNRS,  F-75005 Paris, France}

\author{Eleonora Luppi}
\address{Laboratoire de Chimie Th\'eorique, Sorbonne Universit\'e and CNRS  F-75005 Paris, France}

\author{Elena Degoli}
\address{Dipartimento di Scienze e Metodi dell'Ingegneria, Universit{\`a} di Modena e Reggio Emilia, Via Amendola 2 Padiglione Morselli, I-42122 Reggio Emilia, Italy, \\ 
Centro Interdipartimentale En$\&$Tech, Tecnopolo di Reggio Emilia, I-42124, Reggio Emilia, Italy, \\
Centro S3, Istituto Nanoscienze-Consiglio Nazionale delle Ricerche (CNR-NANO),Via Campi 213/A, 41125 Modena, Italy} 



\begin{abstract}
The interaction of grain boundaries (GBs) with inherent defects and/or impurity elements in multi-crystalline silicon plays a decisive role in their electrical behavior. Strain, depending on the type of grain boundaries and defects, plays important role in these systems. Here, correlation between the structural and electronic properties of $\Sigma$3\{111\} Si-GB in presence of interstitial oxygen impurities have been studied from first-principles framework considering global and local model of strain.
We observed that the distribution of strain along with the number of impurity atoms modify the energetics of the material. 
However, the electronic properties of the considered Si-GB are not particularly affected by the strain and by the oxygen impurities, unless a very high local distortion
induces additional structural defects. 
\end{abstract}


\maketitle

\section{Introduction} 
Multi-crystalline silicon (mc-Si) is the most widely used elemental material on photovoltaic technology in recent times\cite{PV_2009,Chen2011,PV_2017}. Despite the improvement of various aspects, from purification to cell manufacturing including the silicon fabrication processes, current research aims to improve photovoltaic efficiency\cite{solarcell_2020}.
These improvements are primarily dominated by electrical properties of mc-Si systems, which remain limited by several types of defects, particularly the interaction between Si grain boundaries(Si-GBs) and intrinsic point defects or impurity
atoms, as oxygen, carbon, nitrogen, etc.\cite{jcprm_21} and many transition metals\cite{KasinnoJAP13,jap2015_vitaly,OhnoAPL_2016}.
Furthermore, some Si-GBs present residual strain, as detected 
by infrared polariscope or Raman spectroscopy\cite{MSSP2006,JAP2007}, and relation of strain with
longer carrier lifetime\cite{APL2006} has been reported.
Moreover, the distribution of intrinsic strain and its relation to electrical activity on as-grown mc-Si has been probed experimentally in recent times as well\cite{APL2008,JAP2014_GBstrain,JAP2017_GBstrain}. 
Particularly, for metal precipitates correlation between carrier lifetime, stress and precipitate size has been observed\cite{Metal_prc2010}.
Stress not only leads to fracture and breakage during handling and processing, but through the strain produced may also influence
electrical activity and hence affect the performance of furnished solar cells \cite{APL2006,cellefficiency_2011}. Therefore strain play a major role, whether beneficial or not depends on Si-GBs types.

As an intrinsic impurity, large concentration of oxygen is inherent in mc-Si 
\cite{OhnoAPL15}. Oxygen atoms prefer to precipitate at grain boundaries, hence, segregation of oxygen atoms can release the stress, thus lowering the strain energy of the GB \cite{OhnoAPL2017,AM2021}.
However, the mechanisms that control the O segregation are not yet fully understood, due to the diverse and complex nature of different GBs. A simple way to deal with this difficulty is to characterize the local structure at each site around the GBs 
and then to observe the connection between the boundary and these local structures in order to understand how they correlate. 

In this work, we use this strategy and we investigate, from first-principles, the energetics and the electronic properties of Si-GB $\Sigma$3\{111\} with multiple interstitials oxygen atoms both with and without strain. We consider both tensile and compressive strain. For each configuration and type of strain, we analyze the systems revealing factors that mainly influence the system properties.
We compare and discuss different strain varieties that can cause oxygen atoms to interact differently with GB. Actually, understanding how to characterize the structural properties of GBs and their correlation with the electrical activity can be fundamental to understand and control the properties of a device.

\section{Methodology and grain boundary structure}
\label{sec:methogb}
\begin{figure}[t]
\centering
\includegraphics[scale=0.65]{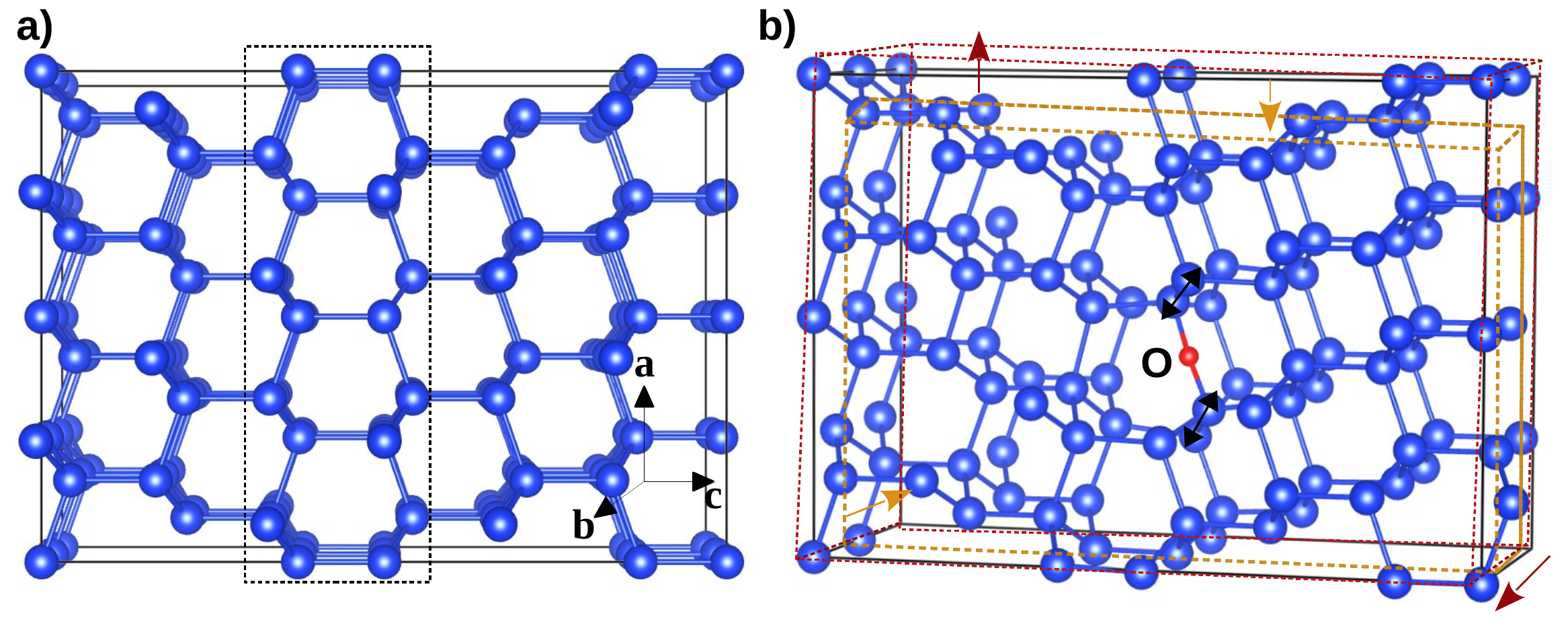}
\caption{a) $\Sigma$3\{111\} Si-GB super-cell: a, b and c are the lattice parameters.
The dotted line shows the GB region, b) schematic representation of global strain: the cell in black-solid line is the unstrained one, while red dotted line and orange dotted line represents the elongated and compressed cells respectively. Local strain is also shown through black arrow in some particular bonds neighboring to the O.}
\label{GB_structure}
\end{figure}

Our model of the $\Sigma$3\{111\} Si-GB consists of two grains of Si forming an interface along the crystallographic plane \{111\} (coincidence site lattice). The two Si grains are misoriented with respect to one another by an angle $\Omega = 60^{\circ}$.  In Figure~\ref{GB_structure}(a) we show the $\Sigma$3\{111\} orthorhombic super-cell ($a$ $\ne$ $b$ $\ne$ $c$ and $\alpha$ = $\beta$ = $\gamma$= $90^{\circ}$) composed of 96 Si atoms. The lattice parameters are $a$=13.30~\AA, $b$=7.68~\AA~and $c$=18.81~\AA. A bi-crystal\cite{jap2015_vitaly} super-cell is adopted to follow periodic boundary conditions. A very regular structure of the $\Sigma$3\{111\} Si-GB, i.e. bond lengths and angles are close to the Si bulk, leads to low formation energy\cite{AM2021}. Concerning O atoms inclusion in the cell, we have tested many possible inequivalent positions of impurities as discussed in\cite{AM2021}. In this paper we focus only on the lowest energy structures (LE).

The calculations were performed using density functional theory (DFT) as implemented in the plane-wave based Vienna Ab-initio Simulation Package (VASP)\cite{Hafner, Kresse}. We employed the generalized gradient approximation PBE for the exchange-correlation functional and projector augmented-wave (PAW) pseudopotentials with a cut-off of 400 eV. K-points sampling within the Monkhorst Pack scheme \cite{Monkhorst} was used for integration of Brillouin-zone together with the linear tetrahedron method including Bl\"ochl corrections\cite{PhysRevB.49.16223}. In particular, we used a k-mesh of 3$\times$3$\times$3 to calculate energy properties of the structures and a k-mesh of 7$\times$7$\times$7 to calculate their density of states (DOS). For the structural optimization, we used as threshold on the forces the value of 10$^{-2}$ eV/\AA{} per atom. All post-processing analysis were carried out using the utilities of VESTA \cite{vesta}.
\section{Results and discussion }
\label{sec:resdisc}
As first step, in the pristine $\Sigma$3\{111\} Si-GB we inserted one by one the O atoms considering different initial configurations and optimizing each structure. Starting from the LE configuration with 1O, different possible position for the second O atoms have been generated and optimized, finally considering the LE configuration with two interstitial O. For three and four oxygen the same approach has been followed. 
For each LE configuration structure with 1O, 2O, 3O and 4O, we calculated the formation energy of oxygen atoms. Moreover, in order to compare all the Si-GBs structures, we considered O interstitial in Si bulk as a reference. In this case, we used a cubic super-cell ($a$ $=$ $b$ $=$ $c$ and $\alpha$ = $\beta$ = $\gamma$= $90^{\circ}$) of 64 atoms with $a$=10.86~\AA.	

We always obtain the O atoms at bond-centered position between two Si atoms and, in the optimized structures, all Si atoms preserve their tetrahedral coordination\cite{OhnoAPL2017,AM2021}. Moreover, the interstitial O atoms can change the structural parameters such as bond lengths and angles, but total energies differ only about 0.01\% irrespective of different positions around GB region.
\begin{figure}[t]
\centering
\includegraphics[scale=0.6]{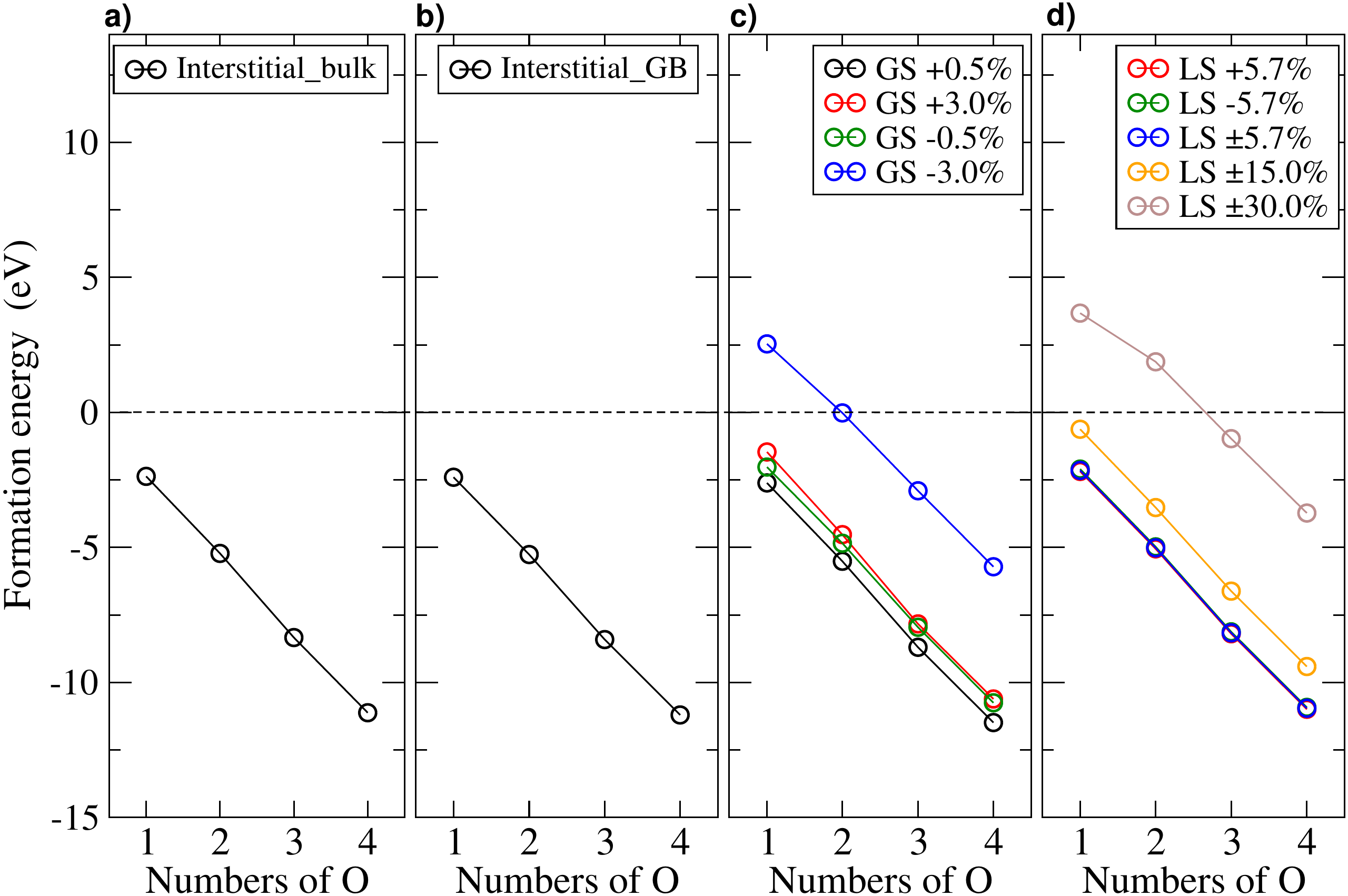}
\caption{Formation energy of interstitial O atoms in (a) Si bulk, (b) $\Sigma$3\{111\} Si-GB, (c) in presence of global strain (GS) and (d) local strain (LS) as mentioned in legends.}
\label{int_energy}
\end{figure} 
 Considering that, according to literature\cite{OhnoAPL_2016,OhnoAPL2017,Ohno_JMC2017}, local distortion plays an important role in the segregation mechanism of O atoms, we have modeled the effect of strain on our structures.
Starting from the LE $\Sigma$3\{111\} Si-GBs along with different numbers of interstitial O atoms ($n$ = 1, 2, 3 and 4), the strain was applied in two different ways, global and local strain, as described in the following:

\begin{itemize}

\item[(1)]Global Strain (GS): in this model, we updated the lattice parameters of the LE O interstitial structure in the x and y directions leaving the z direction fixed. A schematic example of applied GS is shown in Figure~\ref{GB_structure}(b), where an elongation (outward red arrow) and a compression (inward orange arrow) are marked to the $a$ and $b$ lattice parameters. These structures are labeled respectively GS ${+l\%}$ and GS ${-l\%}$ for tensile and compressive strain, where $l$ corresponds to the amount of change with respect to unstrained lattice parameters of LE configuration.
As the elongation or compression is rigidly applied to the whole structure, corresponding changes (elongation or compression) of all bonds in the system are of about the same amount.

\item[(2)]Local Strain (LS)\cite{straindft_2015,AM2021}: in this model, we modified some bond lengths in the interstitial LE structure, creating locally tensile and/or compressive strain in the neighboring Si-Si bonds of O atom(s). In Figure~\ref{GB_structure}(b) we show schematically by black arrows that it is a local change and details will be provided in next sections. Elongation/compression or both are labeled as LS ${+p\%}$, LS ${-p\%}$ and LS ${\pm p\%}$ respectively, where $p$ corresponds to the amount of change with respect to unstrained bond length of the LE configuration. In case of LS ${\pm p\%}$, tensile and compressive strain are applied on two different bonds.
\end{itemize}

To investigate the interaction between the O atoms and the $\Sigma$3\{111\} Si-GB we calculated the impurity formation energy as 

\begin{equation}
 E^{\text{OGB}}_{Imp} = E_{n_{\text{O}}\text{GB}} - E_{\text{GB}}- n_\text{O} \mu_\text{O} 
\label{Gbimp} 
\end{equation}

where $E_{n_{\text{O}}\text{GB}}$ and $E_{\text{GB}}$ are the total energies of the GB including $n_\text{O}$ number of O atoms and of the pristine GB respectively.
$\mu_{\text{O}}$ is the chemical potential of oxygen calculated as the energy per atom of an $O_2$ molecule in vacuum.

This quantity has to be compared with the impurity formation energy ($E^{\text{OB}}_\text{Imp}$) in bulk Si, calculated as 
\begin{equation}
E^{\text{OB}}_{Imp} = E_{n_{\text{O}}\text{B}} - E_{\text{B}}  - n_\text{O} \mu_{\text{O}}
\label{Bulkimp} 
\end{equation}

$E_{n_\text{O}\text{B}}$ and $E_{\text{B}}$ are the total energies of Si bulk containing $n_\text{O}$ number of O atoms and of pristine Si bulk, respectively.

Applying GS and LS methodologies to the LE $\Sigma$3\{111\} Si-GB, we constructed a series of strained structures which differ by the percentage of strain and for which we calculated the impurity formation energy as,

\begin{equation}
 E^{\text{SOGB}}_{Imp} = E_{n_{\text{O}}\text{SGB}} - E_{\text{GB}}- n_\text{O} \mu_\text{O}
\label{sGb} 
\end{equation}

where $E_{n_{\text{O}}\text{SGB}}$ is the total energy of the strained GB including  $n_\text{O}$ number of O atoms.

To keep the effect of the strain during the simulation, it is important not to fully relax the strained structures, which would prefer to relax to their unstrained counterpart\cite{straindft_2015}. Therefore comparing the energy from a self-consistent calculation and the optimized structure with imposing constrain around the O atom to preserve the distortion since they differ by about $3\times10^{-3}$ eV, we resort to the self-consistent calculation\cite{AM2021}.

In Figure~\ref{int_energy} we show the formation energies with different numbers of interstitial O ($n$=1, 2, 3 and 4). In Si bulk [Figure~\ref{int_energy} (a)] and Si-GB [Figure~\ref{int_energy} (b)] energies are almost of similar order. With increasing number of O atoms the lowering of formation energy suggests more stable, hence more favorable systems.  
This trend is preserved in presence of applied strain.
Moreover, for GS [Figure~\ref{int_energy} (c)], compression is much more effective than elongation in destabilizing the system, actually with GS -3.0\% the formation energy become positive and only an increasing number of O atoms is able to stabilize the GB. If the system better sustains a global tensile strain with a large number of O atoms is probably a consequence of the fact that Si and O can mimic the SiO$_2$ configuration, where the lattice parameter is larger with respect to Si.
In case of LS [Figure~\ref{int_energy} (d)], several varieties have been adopted, like different percentage of tensile or compressive strain at both sides of O atoms or a mixed strain, with elongation at one side and compression of the same order of magnitude on the other side ($p$ = $\pm$5.7\%, $\pm$15.0\% and $\pm$30.0\%). Figure~\ref{int_energy} (d) suggests that the system is really stable for LS up to $\pm$15.0\% for different number of interstitial O, only a very large LS (30\%) with a small number of O atoms induce a positive formation energy. 

We then considered how the electronic properties change as a function of strain by focusing only on the systems with $n$=1 and $n$=4 O atoms; actually, we have observed that with $n$=2 and $n$=3 O atoms systems follows similar trend alike formation energy variation.
\begin{figure}[t]
\centering
\includegraphics[scale=0.26]{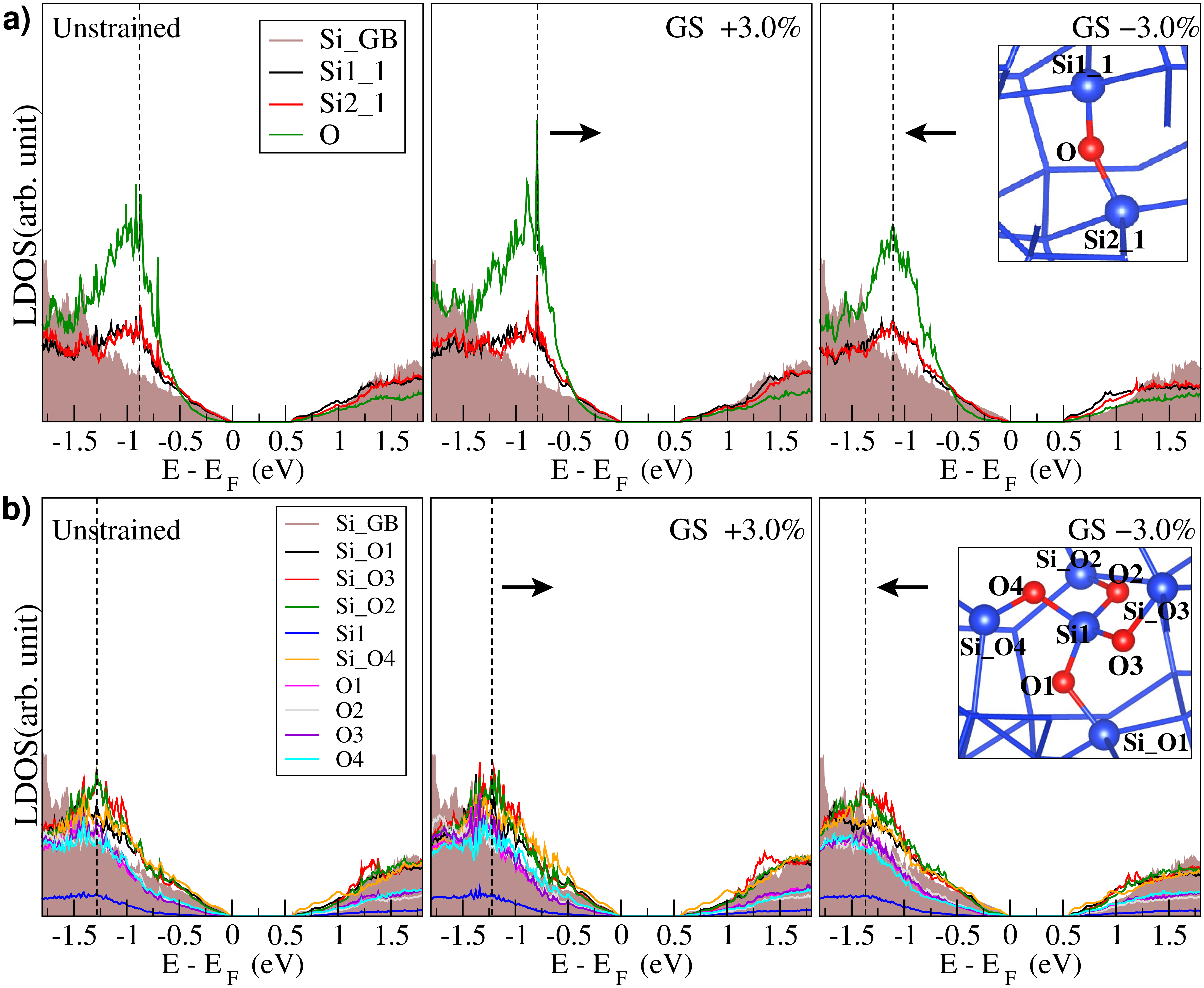}
\caption{ Local density of states (LDOS) in unstrained, under elongation (GS +3.0\%) and compression (GS -3.0\%) for a) one O atom and two Si atoms bonded with O (inset),        
b) four O and all Si atoms connected to O1, O2, O3 and O4 (inset).
Si\_GB refers to a Si atom far from GB region of pristine grain boundary cell.}
\label{strainGS_1O_4O}
\end{figure}

In Figure~\ref{strainGS_1O_4O}(a) we plotted the local density of states (LDOS) for the unstrained GB with 1O (left), strained GB with global tensile strain, GS +3.0\% (center) and global compressive strain, GS -3.0\% (right). Through this comparison it was then possible to clearly analyze how the electronic properties change because of the strain. In order to plot the LDOS we have chosen Si atoms first neighbors of the O and a Si atom (Si\_GB), far from GB region of pristine grain boundary cell. 
Comparing the unstrained LDOS with the tensile GS +3.0\% and the compressive GS -3.0\% one, we observe that there is no particular modification on band edge states or presence of new gap states.
However, due to the elongation (compression) of the Si-O bonds, the delocalization (localization) of the charge density enhances (reduces) the LDOS peak intensity for 1O atom. \\
In Figure~\ref{strainGS_1O_4O}(b) we repeated the same analysis but for the GB with 4O atoms. We obtained the same behavior as for 1O case, although changes are weaker with respect to the unstrained structure. This is consistent with the lowering of the impurity formation energy for the 4O with respect to the 1O atom. Both for 1O and 4O atoms, comparing LDOS of Si around O with that of Si\_GB, we observe that the formation of Si-O bonds introduce new peaks for both Si and O atoms in the energy region around -1.0 eV for 1O and -1.25 eV for 4O. Moreover, in both sets of figures it is clear the effect of strain on the shift of these new peaks actually (as shown by black arrows) while tensile strain shifts peaks higher in energy, nearer to the top of the valence band, compressive strain pushes states to lower energies, deeper in the valence band, with respect to unstrained structures.
\begin{figure}[t]
\centering
\includegraphics[scale=0.18]{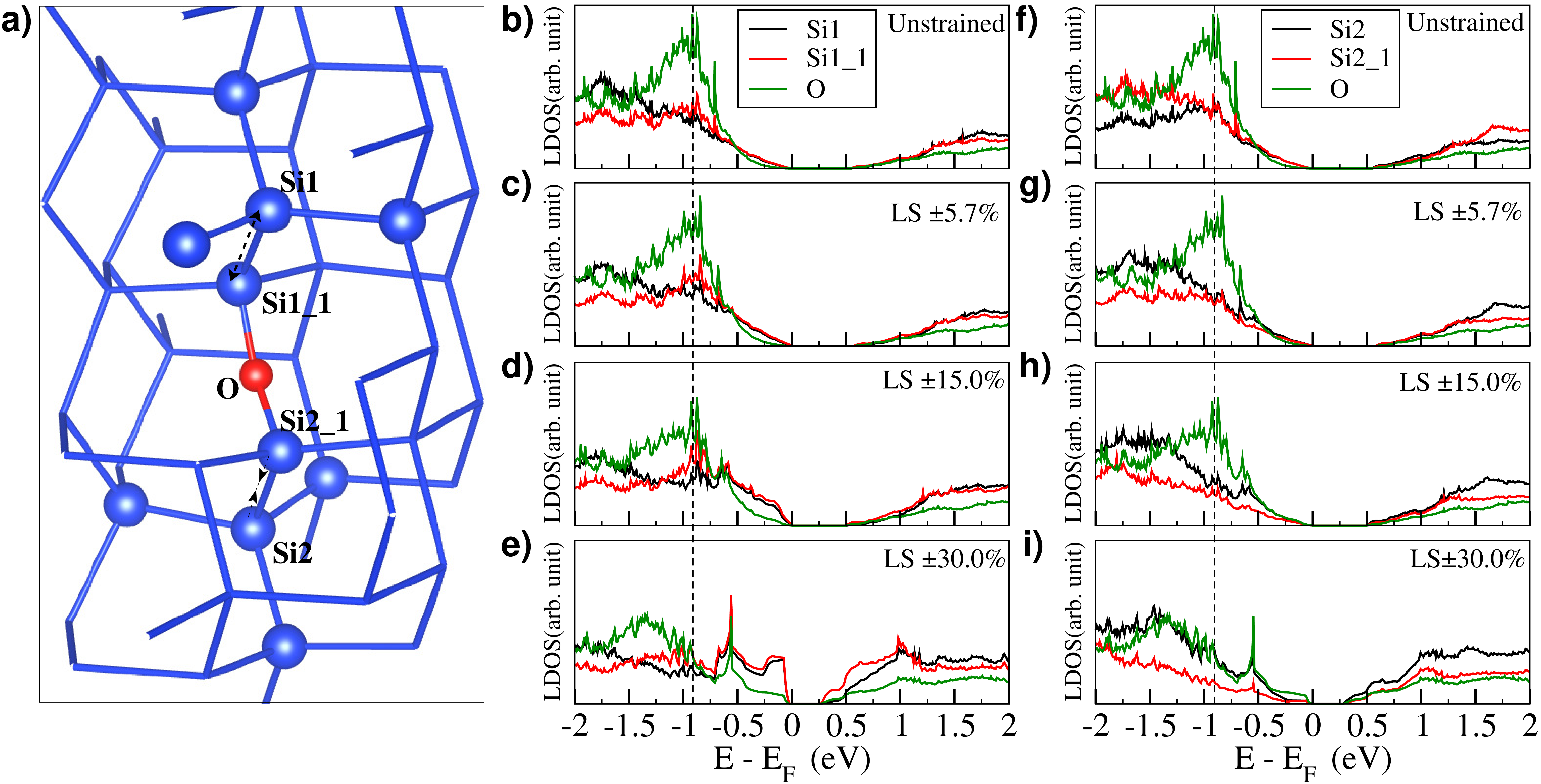}
\caption{a) Local strain structure with one O, Si1 and Si2 are the two sites with reference of which strain has been generated and Si1\_1, Si2\_1 marked the nearest neighbor sites, black dashed arrows are referred to compressed ($->-<-$) or elongated ($<-->$) bonds. Local density of states (LDOS): left panel (b-e) Si1, Si1\_1, O and right panel (f-i) Si2, Si2\_1, O for unstrained and different LS scenarios as mentioned in the inset.}
\label{1O_ls}
\end{figure}

\begin{figure}[t!]
\centering
\includegraphics[scale=0.17]{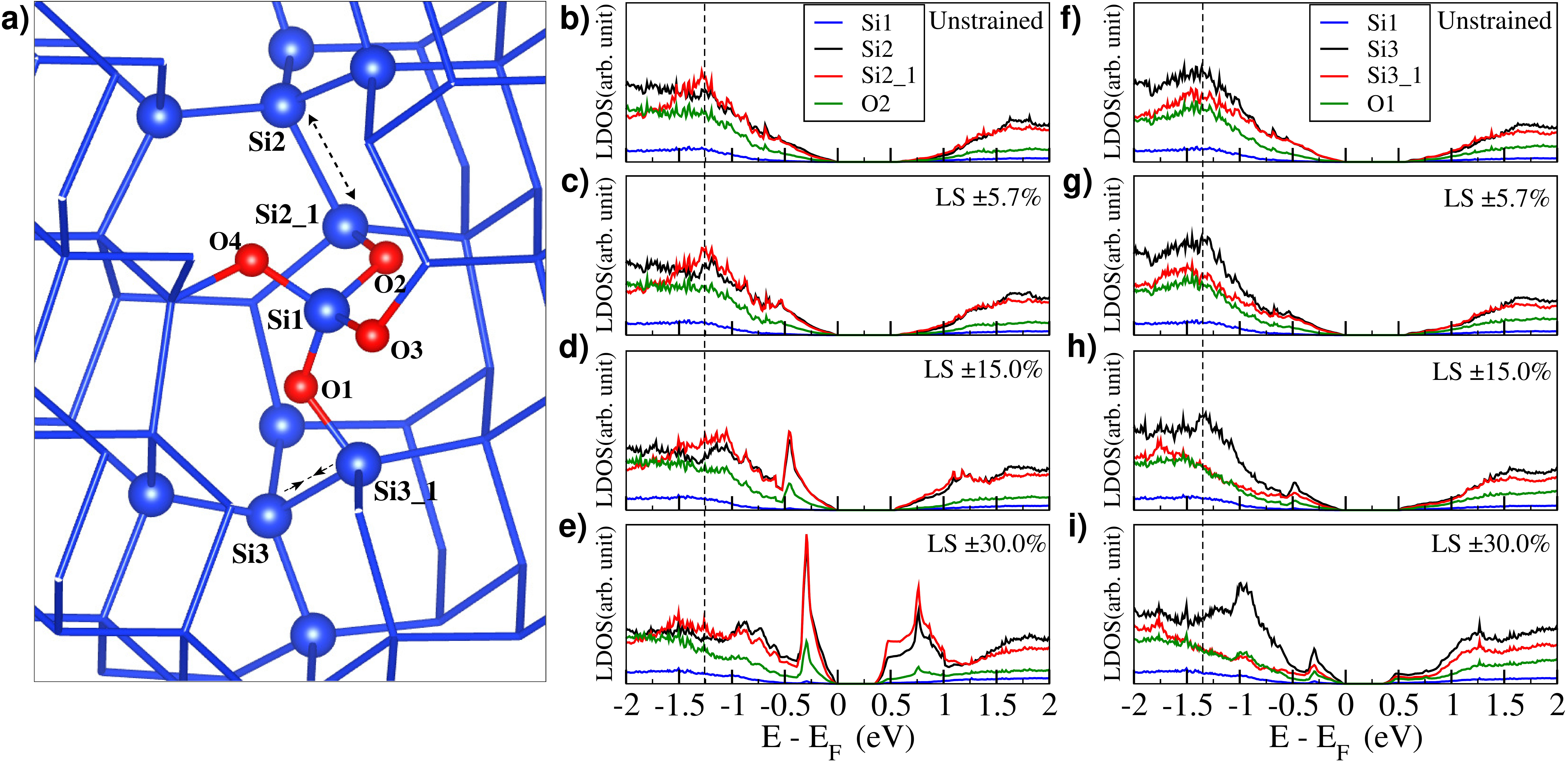}
\caption{a) Local strain structure with four O, Si2 and Si3 are the two sites with reference of which strain has been generated and Si2\_1, Si3\_1 marked the nearest neighbor sites, black dashed arrows are referred to compressed ($->-<-$) or elongated ($<-->$) bonds. Local density of states (LDOS): left panel (b-e) Si1, Si2, Si2\_1, O2 and right panel (f-i) Si1, Si3, Si3\_1, O1 for unstrained and different LS scenarios as mentioned in the inset.} 
\label{4O_ls}
\end{figure}

In Figure~\ref{1O_ls} we show an example of strained Si-GB with 1O (a) and the LDOS (b - i) of the Si atoms and of the O atom as indicated in (a) in the case of an applied local strain. The applied strain along Si1 - Si1\_1 and Si2 - Si2\_1 bonds resort to a deformation of bond length between other three
nearest neighboring sites of Si1 and Si2 as shown with balls in Figure~\ref{1O_ls}(a). 
In each considered structure we have a mixed strain (LS $\pm$5.7\%, LS $\pm$15.0\% and LS $\pm$30.0\%), which means we have elongation along Si1 - Si1\_1 and compression along Si2 - Si2\_1.
Therefore, Figures~\ref{1O_ls}(b - e) show mainly the effect of tensile strain while Figures~\ref{1O_ls}(f - i) show mainly the effect of compressive strain. In both cases it is clear that the electronic properties of this GB are really robust: the smaller distortion ($\pm$5.7\%) has a negligible effect on LDOS and only a strain around or above $\pm$15.0\% is able to alter the electronic configuration of the systems, by changing the peak intensity as well as the energy gap [Table~\ref{gap}]. We would like to note that the +15.0\% tensile strain is a sort of borderline configuration in which the Si1 - Si1\_1 are very weakly bonded (2.687 \AA{}).  
This causes the presence of new structures in the LDOS for Si1 and Si1\_1 at the valence band edge that tend to become defect states. This behavior is enhanced when a +30.0\% strain is introduced as shown in (e). In this case also new conduction band edge states are visible. Compression shown in (g - i) seems to reproduce the same trend but is much less effective because the tetrahedral configuration is preserved and no dangling bonds are formed.\\
For Si-GB with 4O, the LDOS of O1, O2 and the
nearest neighbor Si atoms, i.e. Si1, Si2, Si2\_1 and Si1, Si3, Si3\_1 are shown in Figure~\ref{4O_ls}. 
Here the impact of strain is consistent with 1O, moreover changes in the valence and the conduction band edges, for medium ($\pm$15.0\%) and large ($\pm$30.0\%) strain, show clearly the presence of defect states due to the formation of dangling bond as a consequence of strong distortions. This is mostly evident in Figure~\ref{4O_ls} (e) with high spikes around -0.25 eV and +0.75 eV, where the tensile strain is applied between Si2 and Si2\_1.

\begin{figure}[t!]
\centering
\includegraphics[scale=0.2]{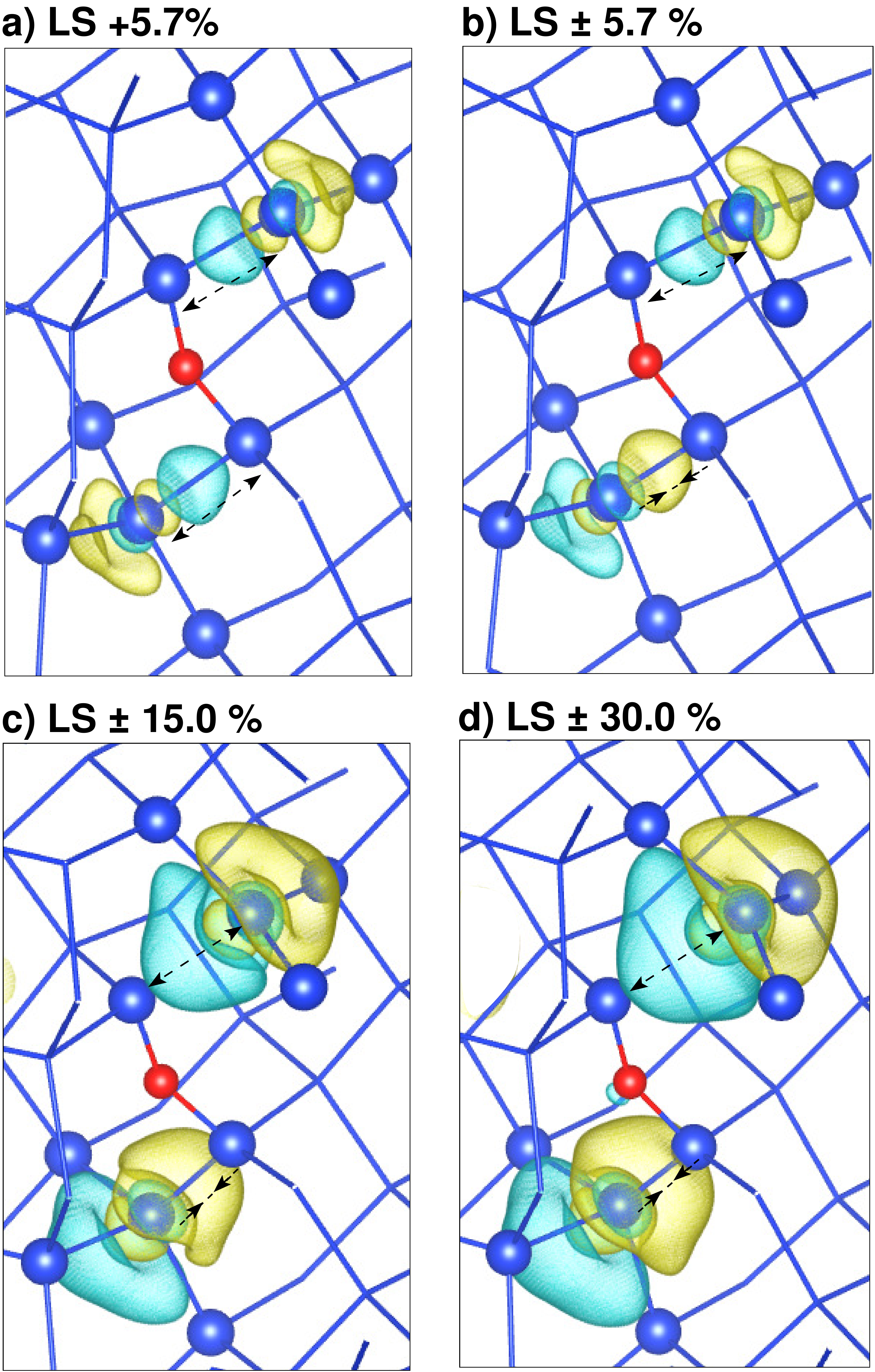}
\caption{Difference of charge density with respect to unstrained structure for 1O with (a) LS +5.7\% , (b) LS $\pm$5.7\% , c) LS $\pm$15.0\%, d) LS $\pm$30.0\%, Yellow (cyan) color corresponds to iso-surface value: +0.007(-0.007). Yellow (cyan) color indicate more (less) amount of charge localization with respect to unstrained structure. Black dashed arrows are referred to compressed ($->-<-$) or elongated ($<-->$) bonds.}
\label{density_dif}
\end{figure}
\begin{figure}[t!]
\centering
\includegraphics[scale=0.56]{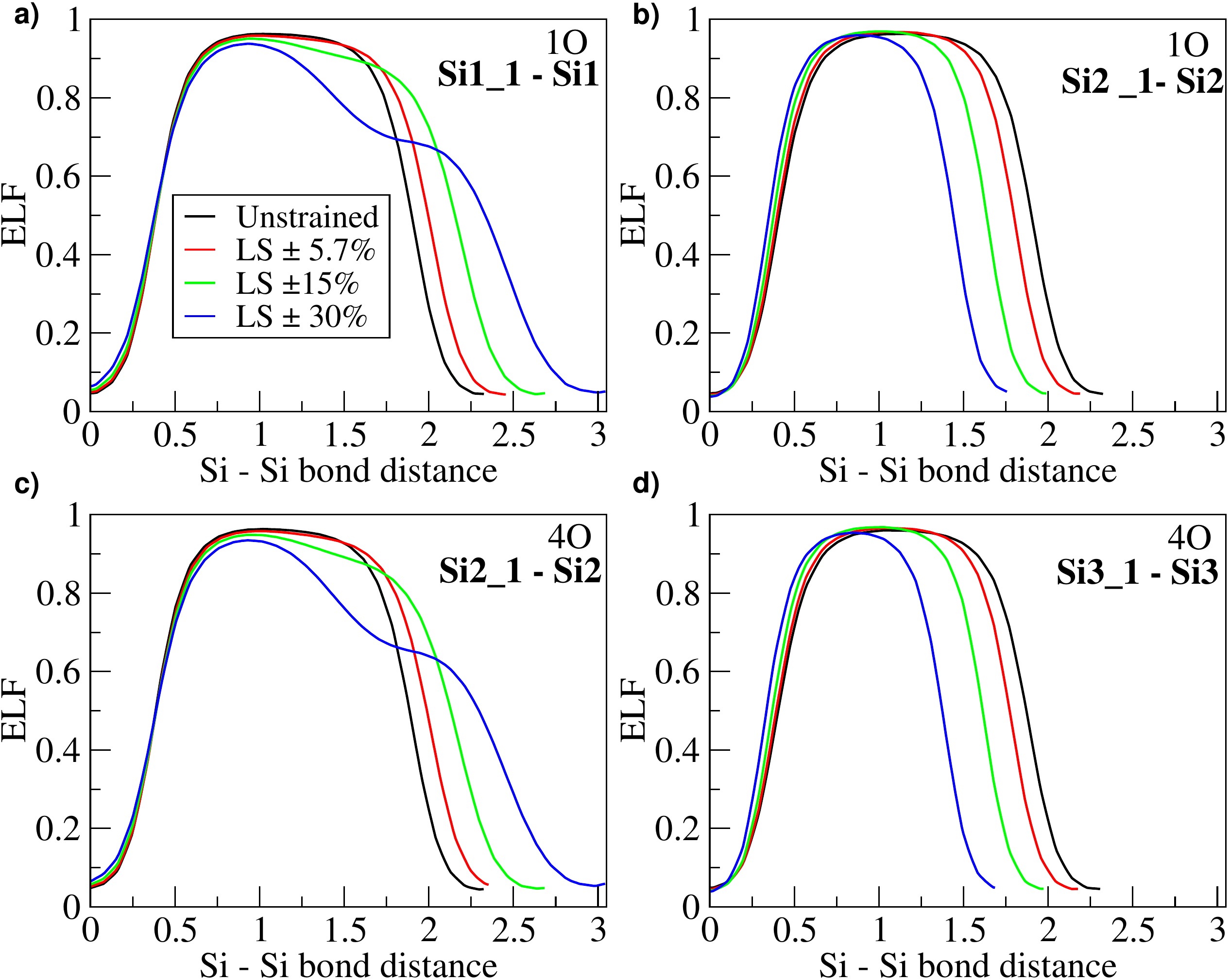}
\caption{ELF profile across the Si - Si strained bond as indicated by the black dashed arrows in Figure~\ref{1O_ls} (a) and \ref{4O_ls}(a) for mixed strain configuration with 1O (a) elongated bond Si1 - Si1\_1, (b) compressed bond Si2 - Si2\_1 and with 4O: (c) elongated bond Si2 - Si2\_1, (d) compressed bond Si3 - Si3\_1 respectively.}
\label{elf_line}
\end{figure}
\begin{table}[h!]
\begin{center}
\begin{tabular}{ |c|c|c| } 
\hline
 Configurations       &1O  &4O
 \\ 
\hline
\hline
Interstitial   &0.551  &0.557 \\ [0.2ex]
\hline
GS +3.0\%      &0.552  &0.545  \\[0.2ex]
\hline
GS +0.5\%      &0.553  &0.557  \\[0.2ex]
\hline
GS $\textendash$0.5\%      &0.548  &0.557  \\[0.2ex]
\hline
GS $\textendash$3.0\%      &0.508  &0.541  \\[0.2ex]
\hline
LS +5.7\%      &0.533  &0.557  \\[0.2ex]
\hline
LS\hspace{+0.2cm}$\textendash$5.7\% &0.544  &0.548  \\[0.2ex]
\hline
LS\hspace{0.04cm}$\pm$ 5.7\%  &0.543 &0.550 \\[0.2ex]
\hline
LS\hspace{0.04cm}$\pm$15.0\% &0.474  &0.527 \\[0.2ex]
\hline
LS\hspace{0.04cm}$\pm$30.0\% &0.283  & 0.362 
\\ 
\hline
\end{tabular}
\end{center}
\caption{Band gap(eV) for Si-GBs with only interstitial and along with global and local strain for ($n$ = 1 and $n$ = 4) oxygen atoms.}
\label{gap}
\end{table}

Band gap values of both GS and LS structures are reported in Table~\ref{gap}. We can observe that band gap changes are more significant in the GB with 1O atom only, where compressive GS red shift the gap while tensile strain is essentially ineffective. In presence of 4O atoms the system is instead really stable.
Actually the relative variation of band gap from unstrained structure is 
maximum $\sim$ 5$\times$ 10$^{-2}$ (eV) for interstitial O atoms ($n$ = 1 and $n$ = 4) when the strain is less than 15.0\% both global or local. Only above this limit, LS starts to have a significant impact on the electronic gap.
In principle, specifically for this indeed stable Si-GB and with interstitial O impurities, vacancy defect would lead to such high local distortion or coordination defect as required to modify the electronic properties\cite{AM2021} significantly.

Moreover, considering that, as previously discussed, LS imposes more inhomogeneous strain field compared to global strain, we have analyzed the charge redistribution through the calculation of the charge density difference (CDD), i.e. the difference between the charge density of the strained system with respect to the unstrained one. 
In Figure~\ref{density_dif} (a - d) CDD have been plotted for 1O with LS +5.7\%, LS $\pm$5.7\%, LS $\pm$15.0\% and LS $\pm$30.0\%. Since, this plot is of similar nature for 4O atoms, we have discussed here only the case with 1O atom. 
In Figure~\ref{density_dif} CDD is locally concentrated on the strained bonds, as expected.
In particular, in Figure~\ref{density_dif} (a, b) CDD for LS +5.7\% and LS $\pm$5.7\% are shown in order to compare tensile only and mixed strain scenario, whereas the effect of different amplitude of LS can be observed from Figure~\ref{density_dif} (b, c and d).
Considering that cyan iso-surfaces correspond to lower charge density while yellow ones correspond to higher charge density with respect to the unstrained system, it is evident that the elongation (compression), shown with dashed arrows in the figure, leads to less (more) charge localization on the strained bonds. 
Actually, for mixed strain [Figure~\ref{density_dif}(b - d)] the yellow iso-surface of CDD in the lower part of the structure where the bond is compressed, and the cyan one in the upper part where the bond is elongated (similarly to the  LS +5.7\% configuration [Figure~\ref{density_dif} (a)]) clearly show the different nature and different effect of the applied LS.
Increasing the percentage of applied tensile and compressive LS [Figure~\ref{density_dif} (b - d)] along Si-Si bonds, the delocalization and localization of charge density respectively, becomes more pronounced. 

Finally, to understand the evolution of bonding structure under LS, we have analyzed the behavior of the electron localization function (ELF)\cite{elf_vasp}. In Figure~\ref{elf_line} ELF profile plots across the strained region, the distorted (elongated/compressed) Si - Si bond in our study, are shown for 1O [Figure~\ref{elf_line} (a, b)] and 4O [Figure~\ref{elf_line} (c, d)] configurations, considering unstrained and mixed LS scenarios.
ELF is a relative measurement of the electron localization and it takes values in the range between 0 and 1. We note that in the PAW method only valence electrons are taken into account in the calculation and for this reason we have the zero value in the ELF profile in the core of the atoms. When the amplitude of ELF is $\sim$1, then electrons are localized in the bonding regions as observed for unstrained configuration (black line), preserving the symmetric linear profile of homopolar Si - Si bond. The same symmetric ELF profile can be observed in the case of compressive strain [Figure~\ref{elf_line} (b, d)], where the tetrahedral coordination is preserved and the reduction of the Si-Si bond length results in a stronger electron localization.
Concerning the tensile strain, [Figure~\ref{elf_line} (a, c) show a different ELF profile that progressively loses its symmetry with increasing percentage of strain. Actually for strong elongation the bond begins to dissolve and for the maximum LS $\pm$30.0\% the electron, no more involved in a bond, tend to localize more towards one Si atom (Si1\_1 for 1O and Si2\_1 for 4O) than the other (Si1 for 1O and Si2 for 4O).
The ELF analysis confirms and clarifies the effect of different strain categories applied to the considered GBs on the local charge redistribution.

\section{Conclusions}
In conclusion, formation energy depends on geometrical distortions, either distributed homogeneously or inhomogeneously under GS or LS model, and on the number of impurity precipitates.
We observe that the electronic properties are almost unchanged in presence of a small strain and formation energies decrease with increasing number of O atoms. For GS, this can be explained by the fact that silicon atoms preserves their tetrahedral coordination. Moreover, global tensile strain resulted to be relatively preferable for the GB stability due to the fact that under tensile strain silicon and oxygen can mimic the structure of SiO$_2$ which in fact has a larger lattice parameter than Si. Considering instead the effect of local distortion [Figure~\ref{1O_ls},      
 \ref{4O_ls}] while small strain is essentially ineffective, strong tension above +15\%, as the one eventually produced by vacancies, is more efficient than compression in changing the electronic properties of the GBs. Actually a large tension can induce dangling bonds formation and the presence of defect states in the gap.
Moreover, through the analysis of the charge density and of the electron localization function we have demonstrated that while the bond elongation reduces the charge density localized on the bond, the compression increases it.
The fact that, a strain below 15.0\% does not directly affect the electrical activity of this particular Si-GB, thus even with oxygen precipitates, preserving the robustness of electronic properties, could be a useful understanding in modeling devices with such samples. However, the electronic properties and the effect of interstitial oxygen(s) and applied strain may vary with the different types of Si-GB geometry and this thing need to be analyzed in more detail.

\medskip
\section{Acknowledgements}
We would like to thank the University of Modena and Reggio Emilia for the financial support (FAR dipartimentale 2020) and Centro Interdipartimentale En$\&$Tech, as well as the CINECA HPC facility for the approved ISCRA C project SiGB-NMI (IsC86\_SiGB-NMI).

\medskip

\end{document}